\documentclass[sigconf]{acmart}

\AtBeginDocument{%
  }

\acmYear{2025}
\setcopyright{cc}
\acmConference[PEARC '25]{Practice and Experience in Advanced Research Computing}{July 20--24, 2025}{Columbus, OH, USA}
\acmBooktitle{Practice and Experience in Advanced Research Computing (PEARC '25), July 20--24, 2025, Columbus, OH, USA}
\acmDOI{10.1145/3708035.3736049}
\acmISBN{979-8-4007-1398-9/2025/07}

\acmISBN{}

\usepackage{acronym}
\acrodef{AI}{artificial intelligence}
\acrodef{DL}{deep learning}
\acrodef{CNNs}{convolutional neural networks}
\acrodef{RNN}{recurrent neural network}
\acrodef{GNN}{graph neural network}
\acrodef{GNNs}{graph neural networks}
\acrodef{GDL}{geometric deep learning methods}
\acrodef{NN}{neural network}
\acrodef{ML}{machine learning}
\acrodef{FML}{Fast Machine Learning}
\acrodef{SM}{standard model}
\acrodef{CMS}{Compact Muon Solenoid}
\acrodef{QCD}{quantum chromodynamics}
\acrodef{LHC}{Large Hadron Collider}
\acrodef{HL-LHC}{High-Luminosity LHC}
\acrodef{QFT}{quantum field theory}
\acrodef{IRC}{infrared and collinear}
\acrodef{EFN}{energy flow network}
\acrodef{EMD}{earth mover's distance}
\acrodef{HPC}{high-performance computing}
\acrodef{A3D3}{Accelerated AI Algorithms for Data-Driven Discovery}
\acrodef{ASAML}{Advancing Science with Accelerated Machine Learning}
\acrodef{HDR}{Harnessing the Data Revolution}
\acrodef{LIGO}{the Laser Interferometer Gravitational-Wave Observatory}
\acrodef{MMA}{multi-messenger astrophysics}
\acrodef{HEP}{high energy physics}
\acrodef{HLT}{high-level trigger}
\acrodef{L1T}{level-1 trigger}
\acrodef{BSM}{beyond the SM}
\acrodef{GW}{gravitational wave}
\acrodef{DUNE}{Deep Underground Neutrino Experiment}
\acrodef{LAr}{liquid argon}
\acrodef{TPC}{time-projection chamber}
\acrodef{ZTF}{Zwicky Transient Facility}
\acrodef{FPGA}{field-programmable gate array}
\acrodef{GPU}{graphics processing unit}
\acrodef{IPU}{Intelligence Processing Unit}
\acrodef{TPU}{tensor processing unit}
\acrodef{HLS}{high-level synthesis}
\acrodef{ASIC}{application-specific integrated circuit}
\acrodef{CUDA}{Compute Unified Device Architecture}
\acrodef{SONIC}{Services for Optimized Network Inference on Coprocessors}
\acrodef{L1DS}{level-1 data scouting}
\acrodef{pp}{proton-proton}
\acrodef{URM}{underrepresented minority}
\acrodef{MLP}{multilayer perceptron}
\acrodef{WCC}{weekly-connected-component}
\acrodef{ACTS}{A Common Tracking Software}
\acrodef{L1}{level-1}
\acrodef{MC}{Monte Carlo}
\acrodef{NRP}{National Research Platform}
\acrodef{ATLAS}{A Toroidal LHC Apparatus}
\acrodef{IceCube}{the IceCube Neutrino Observatory}
\acrodef{KEDA}{Kubernetes Event-Driven Autoscaling}
\acrodef{kind}{Kubernetes IN Docker}
\acrodef{CVMFS}{CernVM Filesystem}
\acrodef{KAGRA}{the Kamioka Gravitational Wave Detector}
\acrodef{CNCF}{Cloud Native Computing Foundation}
\acrodef{MIG}{Multi-Instance GPU}
\acrodef{NSF}{National Science Foundation}

\begin{document}

\title[SuperSONIC]{SuperSONIC: Cloud-Native Infrastructure for ML Inferencing}

\author{Dmitry Kondratyev}
\affiliation{%
  \institution{Purdue University}
  \city{West Lafayette}
  \state{Indiana}
  \country{USA}
}
\email{dkondra@purdue.edu}
\orcid{0000-0002-7874-2480}

\author{Benedikt Riedel}
\affiliation{%
  \institution{University of Wisconsin–Madison}
  \city{Madison}
  \state{Wisconsin}
  \country{USA}
}
\email{briedel@icecube.wisc.edu}
\orcid{0000-0002-9524-8943}

\author{Yuan-Tang Chou}
\affiliation{%
  \institution{University of Washington}
  \city{Seattle}
  \state{Washington}
  \country{USA}
}
\email{ytchou@uw.edu}
\orcid{0000-0002-2204-5731}

\author{Miles Cochran-Branson}
\affiliation{%
  \institution{University of Washington}
  \city{Seattle}
  \state{Washington}
  \country{USA}
}
\email{milescb@uw.edu}
\orcid{0000-0003-1020-1108}

\author{Noah Paladino}
\affiliation{%
  \institution{Massachusetts Institute of Technology}
  \city{Cambridge}
  \state{Massachusetts}
  \country{USA}
}
\email{npaladin@mit.edu}
\orcid{0000-0003-1225-537X}

\author{David Schultz}
\affiliation{%
  \institution{University of Wisconsin–Madison}
  \city{Madison}
  \state{Wisconsin}
  \country{USA}
}
\email{dschultz@icecube.wisc.edu}
\orcid{0000-0002-7841-3919}


\author{Mia Liu}
\affiliation{%
  \institution{Purdue University}
  \city{West Lafayette}
  \state{Indiana}
  \country{USA}
}
\email{liu3173@purdue.edu}
\orcid{0000-0001-9012-395X}

\author{Javier Duarte}
\affiliation{%
  \institution{University of California San Diego}
  \city{La Jolla}
  \state{California}
  \country{USA}
}
\email{jduarte@ucsd.edu}
\orcid{0000-0002-5076-7096}

\author{Philip Harris}
\affiliation{%
  \institution{Massachusetts Institute of Technology}
  \city{Cambridge}
  \state{Massachusetts}
  \country{USA}
}
\email{pcharris@mit.edu}
\orcid{0000-0001-8189-3741}

\author{Shih-Chieh Hsu}
\affiliation{%
  \institution{University of Washington}
  \city{Seattle}
  \state{Washington}
  \country{USA}
}
\email{schsu@uw.edu}
\orcid{0000-0001-6214-8500}

\renewcommand{\shortauthors}{Kondratyev, et al.}

\begin{abstract}
The increasing computational demand from growing data rates and complex \ac{ML} algorithms in large-scale scientific experiments has driven the adoption of the \ac{SONIC} approach. \ac{SONIC} accelerates \ac{ML} inference by offloading it to local or remote coprocessors to optimize resource utilization. Leveraging its portability to different types of coprocessors, \ac{SONIC} enhances data processing and model deployment efficiency for cutting-edge research in \ac{HEP} and \ac{MMA}. We developed the SuperSONIC project, a scalable server infrastructure for \ac{SONIC}, enabling the deployment of computationally intensive tasks to Kubernetes clusters equipped with \acp{GPU}. Using NVIDIA Triton Inference Server, SuperSONIC decouples client workflows from server infrastructure, standardizing communication, optimizing throughput, load balancing, and monitoring. SuperSONIC has been successfully deployed for the CMS and ATLAS experiments at the CERN \ac{LHC}, \ac{IceCube}, and \ac{LIGO} and tested on Kubernetes clusters at Purdue University, the \ac{NRP}, and the University of Chicago. SuperSONIC addresses the challenges of the Cloud-native era by providing a reusable, configurable framework that enhances the efficiency of accelerator-based inference deployment across diverse scientific domains and industries.

\end{abstract}



\maketitle

\acresetall

\section{Introduction}

The impact of \ac{AI} and \ac{ML} on society in general and science in particular has been likened to the introduction of the steam engine or the internet by altering how we interact with machines and ``do'' science.
Within a few years \ac{AI}, and in particular \ac{ML}, has revolutionized and accelerated scientific discovery across multiple fields to such a magnitude that the 2024 Nobel Prizes in Physics and Chemistry were awarded for fundamental work on \ac{AI} (Physics) and new discoveries made using \ac{AI} (Chemistry)~\cite{nobelprize24}. 

\ac{AI}/\ac{ML} requires a heterogeneous computing paradigm, using traditional (CPU) and specialized (generally \acp{GPU}) hardware for both training and evaluating (``inferencing'') the model. \ac{AI} training has caught the headlines requiring ever larger heterogeneous clusters growing to over 100,000 \acp{GPU}~\cite{grokgpucluster} with plans reaching up to 1,000,000 \acp{GPU}~\cite{grokgpucluster1M}. By comparison, inferencing an \ac{AI}/\ac{ML} model can be done with a single lower-performance GPU. This metric, however, is misleading. A given \ac{AI}/\ac{ML} model generally gets inferenced so often that 90\% of the lifetime and computing cost of an \ac{AI}/\ac{ML} model are estimated to be spent on inferencing alone~\cite{ibminference,awsinference}. 

Large-scale and data-intensive experiments, such as \ac{IceCube}, the \ac{LHC}'s \ac{CMS} and \ac{ATLAS}, and \ac{LIGO} experiments, all of which are represented here, are adding and increasingly relying on \ac{AI}/\ac{ML} throughout the experiment ranging from data acquisition to data processing and final analysis~\cite{duarte2024novelmachinelearningapplications}. This trend is expected to continue as these experiments expand their science reach with \ac{AI}-enhanced data analyses, detector upgrades, e.g. \ac{HL-LHC}~\cite{hllhc} and \ac{IceCube} Upgrade~\cite{ishihara2019icecubeupgradedesign}, and new experiments coming online, e.g. \ac{KAGRA}~\cite{akutsu2020overviewkagradetectordesign}. 

Given the limited computational and human resources available to scientists, the growing use of \ac{AI} in experiments, and the potentially high cost of \ac{AI} inferencing, optimizing \ac{AI}/\ac{ML} deployment across the national cyberinfrastructure is crucial.
The science community is taking several different approaches depending on the needs and requirements of the respective science driver. For high-throughput environments, such as the \ac{L1T} for the \ac{LHC} experiments~\cite{duarte2024novelmachinelearningapplications}, the focus has been on integrating \acp{FPGA} for \ac{AI} inference to increase overall throughput compared to \acp{GPU}. In less demanding environments, such as higher-level data processing, the development has not been as focused. Most scientists still co-locate heterogeneous compute resources for their \ac{AI}/\ac{ML} inference needs. However, co-location of heterogeneous resources limits the ability for the system to adapt to changes in the software stack that would require a different ``mix'' of resources (e.g., CPU/\ac{GPU}), leading to significant inefficiencies in resource utilization.  

To address this issue, \ac{SONIC}~\cite{Duarte:2019fta,Krupa:2020bwg,Rankin:2020usv,Wang:2020fjr,CMS:2024twn,Zhao:2025daw} were developed based on the ``as-a-service'' model in which the \ac{AI}/\ac{ML} compute is offloaded to remote accelerated resources. 
The idea of SONIC originated from the private sector, where the local hardware available is typically limited, e.g., smartphones, or insufficient to run large-scale models, e.g., ChatGPT~\cite{chatgpt}.

As the SONIC approach was being implemented in \ac{ATLAS}, \ac{CMS}, \ac{IceCube}, and \ac{LIGO} experiments, it became apparent that, despite the differences in client workflows, the server infrastructure requirements are largely the same. Existing industry-grade inference platforms such as KServe~\cite{kserve}, vLLM~\cite{vllm}, and NVIDIA Triton~\cite{triton} could not be used ``off the shelf'', as their functionality did not meet all the needs of scientific workflows. 
Faced with the same challenge, the aforementioned experiments have joined forces under the \ac{NSF} funded \ac{HDR} \ac{A3D3} Institute to develop SuperSONIC~\cite{supersonic}, a common cloud-native inference framework designed to deploy \ac{ML}-inference-as-a-service on national cyberinfrastructure.

\section{Design}

SuperSONIC is designed as a scalable and portable system capable of managing inference workloads across diverse environments, from tiny GitHub Actions workers to small local clusters to large distributed \ac{HPC} centers. Its modular architecture relies on established industry-standard tools, with a preference given to open-source projects with guaranteed long-term maintainability and community support, such as the ``graduated'' projects of the \ac{CNCF}~\cite{cncf}.

\begin{figure*}[h]
    \centering
    \includegraphics[width=\linewidth]{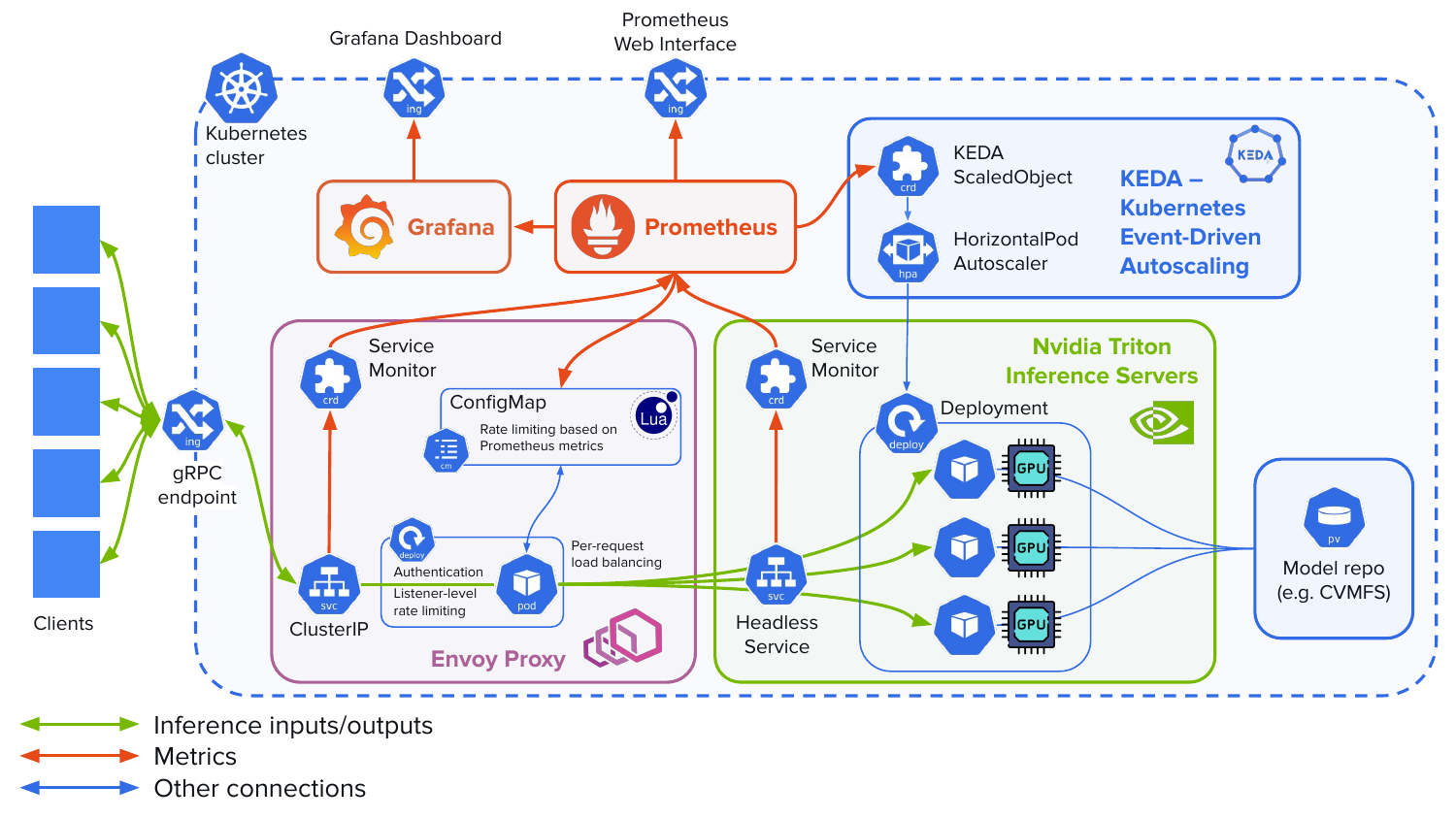}
    \caption{SuperSONIC architecture: client nodes (left) running specialized experimental software outsource the heterogeneous, often AI/ML, component of the software stack to a SuperSONIC server deployed on a Kubernetes cluster. The server handles dynamic GPU provisioning, optimized load balancing, and other operational complexities, and exposes only a single gRPC endpoint for inference requests.}
    \label{fig:diagram}
\end{figure*}

SuperSONIC leverages Kubernetes~\cite{kubernetes} to deploy its components as microservices, ensuring seamless workload orchestration and fault tolerance while abstracting infrastructure complexities into a simple, declarative configuration. To streamline installation and version control, it is distributed as a Helm chart~\cite{helm}.
Figure~\ref{fig:diagram} illustrates the core components of the SuperSONIC architecture, which are described below in more detail.

\subsection{Inference Servers}

SuperSONIC is built around NVIDIA Triton Inference Server~\cite{triton}, which provides a robust and efficient way to serve machine learning models at scale and accelerate their evaluation using \acp{GPU}.  At present, only NVIDIA \acp{GPU} are supported, with plans to extend compatibility to other \ac{GPU} vendors in the future through the integration of PyTriton~\cite{pytriton}.

Triton loads models from \textit{model repositories}, which can be configured based on specific infrastructure requirements. Available options include Kubernetes persistent storage with support for various backends, the distributed \ac{CVMFS}, as well as mounted NFS volumes.

\subsection{Proxy}

A critical component of SuperSONIC is the Envoy Proxy~\cite{envoy}, which acts as the gateway between clients and inference servers. It plays a key role in optimizing server performance, managing traffic, and preventing overloads.

\begin{itemize}
    \item \textbf{Load balancing} distributes incoming requests across multiple Triton instances using predefined algorithms such as round robin.
    \item \textbf{Rate limiting} regulates server load based on the number of client connections or on an arbitrary external metric.
    \item \textbf{Token-based authentication} secures client endpoints, preventing unauthorized access and mitigating potential misuse.
\end{itemize}

\subsection{Monitoring Tools}
SuperSONIC collects various metrics from its components to monitor system health and performance, as well as to provide figures of merit for load-based rate limiting and autoscaling.
Among the key collected metrics are the inference rate for each evaluated model, a breakdown of total request latency by source, and GPU engine and memory utilization.
Metrics collection is implemented via Prometheus~\cite{prometheus}, with the option to deploy a custom Prometheus instance or connect to an existing one. The metrics are visualized using Grafana~\cite{grafana}, and a pre-configured Grafana dashboard is automatically installed with the SuperSONIC deployment.
For deeper insights, tracing via OpenTelemetry~\cite{opentelemetry} and Grafana Tempo~\cite{tempo} enables a more detailed analysis of inference request flows and performance bottlenecks.

\subsection{Autoscaling}

Given the relative scarcity of \ac{GPU} resources, it is crucial to ensure their efficient utilization in the inference pipelines. An overloaded inference server diminishes the benefits of \ac{GPU} acceleration, as incoming requests are forced to wait in a queue before getting evaluated.  On the other hand, underutilized \acp{GPU} consume valuable resources that could be allocated elsewhere and lead to unnecessary cost overheads.

To address this, SuperSONIC employs \ac{KEDA}~\cite{keda} to dynamically adjust resource allocation based on workload demand. \ac{KEDA} is configured to launch additional Triton instances when a user-defined metric exceeds a given threshold and, conversely, to shut down servers when the metric value falls below the threshold. The default scaling metric is defined as the average request queue latency across Triton servers.

\section{Deployment and Testing}

SuperSONIC has been tested in a variety of scenarios to showcase its versatility, portability, and scalability.
Notably, remote \ac{GPU} acceleration was demonstrated for \ac{GNNs} and transformer architectures in the CMS experiment, \ac{CNNs} in the \ac{IceCube} and \ac{LIGO} experiments, and non-ML tracking algorithms in the ATLAS experiment. Despite significant differences on the client side, different workflows were shown to benefit from a common server-side implementation, highlighting the versatility of the ``inference-as-a-service'' approach.

A complete SuperSONIC deployment, along with a generic client workflow, was shown to fit inside a single GitHub Actions~\cite{github-actions} worker (4 CPU cores, 16 GB memory) utilizing a small \ac{kind}~\cite{kind} cluster. On the other hand, a SuperSONIC deployment at the \ac{NRP}~\cite{nrp1, nrp2} was tested with as many as 100 GPU-enabled Triton servers. Such flexibility makes SuperSONIC suitable for a wide range of applications in science.

The SuperSONIC package was deployed with minimal differences on the Geddes~\cite{geddes} and Anvil~\cite{anvil} clusters at Purdue University, at the \ac{NRP}, and on the \ac{ATLAS} Analysis Facility~\cite{uchicago-af} at the University of Chicago, proving that the software is highly adaptable to different computing environments.

\section{Performance}
The effectiveness of SuperSONIC becomes most evident when inference workload intensity varies over time. In such scenarios, static GPU allocation is inefficient: it either suffers high latency when GPUs are overloaded or wastes resources under light load. Load-based autoscaling in SuperSONIC, on the other hand, enables dynamic adjustment of GPU allocation in response to changing workload.
To demonstrate this behavior, a synthetic workflow was constructed using NVIDIA Triton Performance Analyzer~\cite{perf-analyzer} clients that evaluate the ParticleNet~\cite{particlenet} model -- a \ac{GNN} used in the \ac{CMS} experiment. The batch size was configured such that a single NVIDIA T4 GPU could sustain the inference load from one client but not from ten clients running in parallel.

\begin{figure}
    \centering
    \includegraphics[width=\linewidth]{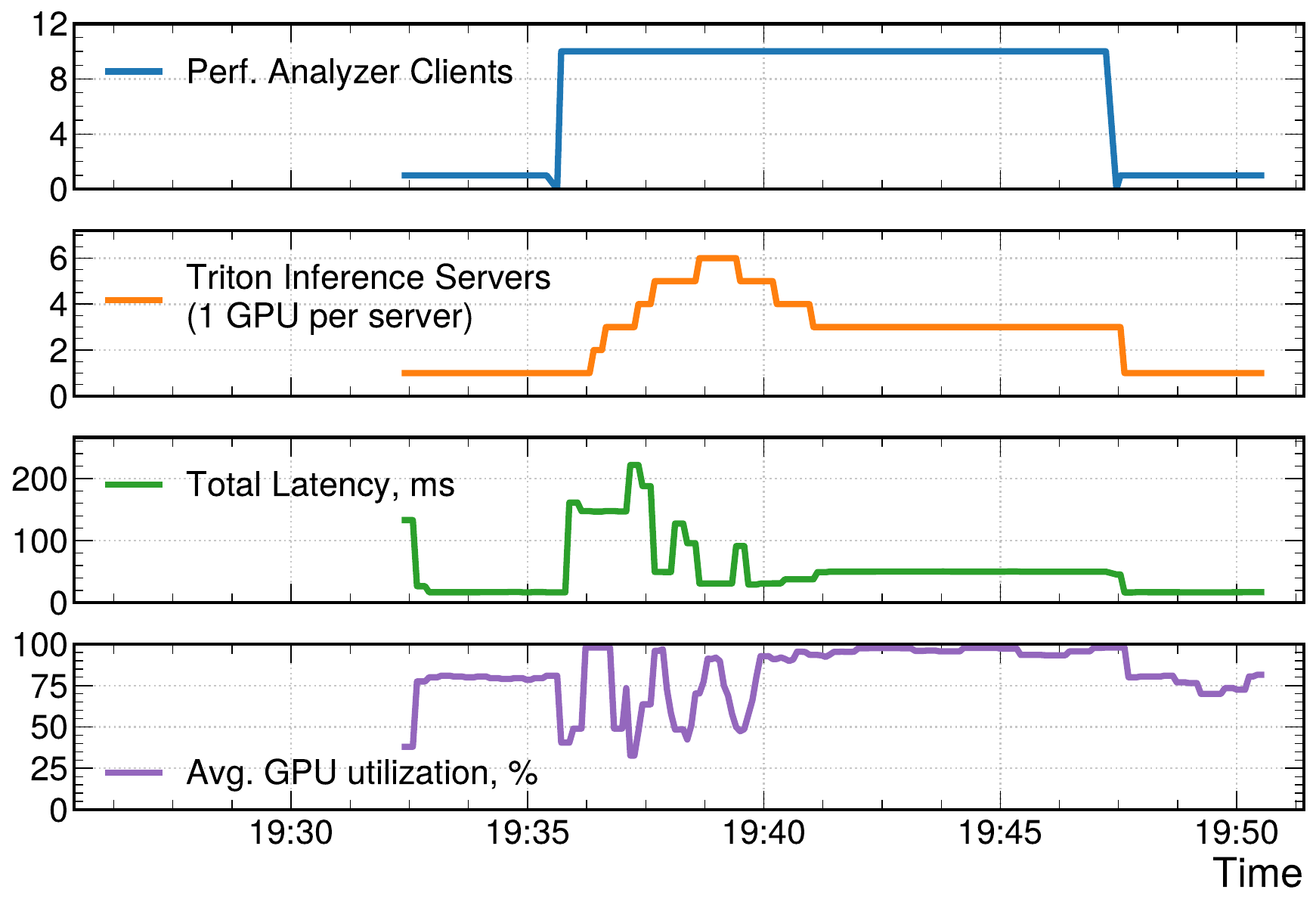}
    \caption{Load-based autoscaling in SuperSONIC: the GPU server count (orange) adjusts in response to spikes in latency (green) caused by increased inference load (blue).}
    \label{fig:autoscaling}
\end{figure}

\begin{figure}
    \centering
    \includegraphics[width=\linewidth]{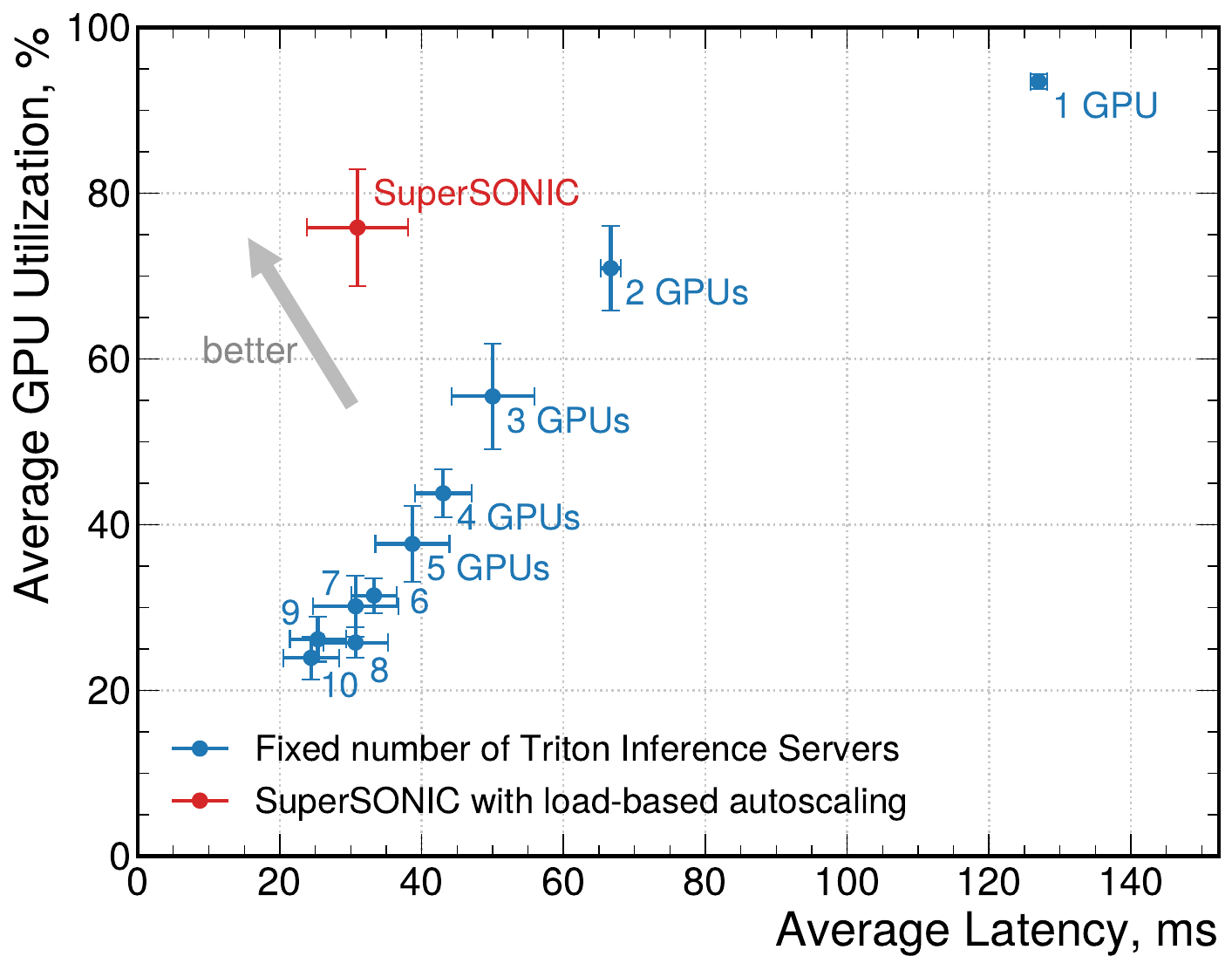}
    \caption{Average GPU utilization and latency for a test workflow with an inference load that varies over time. Dynamic GPU provisioning with SuperSONIC (red) outperforms setups with fixed GPU count (blue).}
    \label{fig:util_vs_latency}
\end{figure}

Figure~\ref{fig:autoscaling} illustrates SuperSONIC's autoscaling behavior under varying inference load. Initially, the light load from a single client is handled by a single GPU server. When the number of clients is increased to ten, the latency spikes, triggering the number of GPU servers to scale up. The number of GPUs eventually settles on the value that provides the optimal trade-off between GPU utilization and latency. Finally, as the load is decreased back to one client, SuperSONIC releases unneeded GPUs.
The same workflow ($1\rightarrow 10 \rightarrow 1$ clients) was executed against deployments with a static number of GPU-enabled Triton Inference Servers. Figure~\ref{fig:util_vs_latency} shows the average latency and GPU utilization measured over the duration of each test. Compared to static configurations, the dynamic GPU allocation in SuperSONIC significantly improves both metrics. The trade-off between latency and GPU utilization can be further adjusted by tuning the responsiveness of the autoscaler, as well as the metric used as its trigger.

\section{Conclusion}

The rapid adoption of \ac{AI}/\ac{ML} across scientific domains highlights the growing role of hardware accelerators and the need for optimized and scalable inference solutions. The \ac{SONIC} design paradigm, based on the concept of decoupling server infrastructure from client workflows, allows for efficient utilization of heterogeneous resources. Specifically, it addresses our inability to predict the optimal CPU-to-heterogeneous ratio for each \ac{AI}/\ac{ML} application, instead offering a dynamic optimization approach. Such an approach is particularly suited to computing models of large scientific experiments, where heterogeneous resources are unevenly distributed between computing sites, and algorithm developments are frequently changing the amount of heterogeneous resources needed. The SuperSONIC project, conceived within the \ac{NSF} \ac{HDR} \ac{A3D3} Institute and co-developed by multiple scientific experiments, implements the \ac{SONIC} paradigm as a cloud-native inference platform based on state-of-the-art industry tools. The compatibility of SuperSONIC with various computing workflows and model architectures, combined with its portability across diverse computing environments, makes it a useful tool in scientific research to ensure optimal heterogeneous compute workflows in any scientific domain. 

\begin{acks}

D. Kondratyev and M. Liu were supported by the U.S. CMS Software and Computing Operations Program under the U.S. CMS HL-LHC R\&D Initiative. Y. Chou, M. Cochran-Branson, J. Duarte, P. Harris, S. Hsu, M. Liu, B. Riedel, and D. Schultz were supported by \ac{NSF} grant No. PHY-2117997. This work was supported in part by \ac{NSF} awards CNS-1730158, ACI-1540112, ACI-1541349, OAC-1826967, OPP-2042807, OAC-2112167, CNS-2100237, CNS-2120019.

\end{acks}

\balance

\bibliographystyle{ACM-Reference-Format}


\end{document}